\documentstyle[twoside]{article}

\catcode`\@=11
\long\def\@makefntext#1{
\protect\noindent \hbox to 3.2pt {\hskip-.9pt  
$^{{\eightrm\@thefnmark}}$\hfil}#1\hfill}               %CAN BE USED 

\def\@makefnmark{\hbox to 0pt{$^{\@thefnmark}$\hss}}    %ORIGINAL 
        
\def\ps@myheadings{\let\@mkboth\@gobbletwo
\def\@oddhead{\hbox{}
\rightmark\hfil\eightrm\thepage}   
\def\@oddfoot{}\def\@evenhead{\eightrm\thepage\hfil
\leftmark\hbox{}}\def\@evenfoot{}
\def\sectionmark##1{}\def\subsectionmark##1{}}

\oddsidemargin=\evensidemargin
\newcounter{sectionc}\newcounter{subsectionc}\newcounter{subsubsectionc}
\renewcommand{\section}[1] {\vspace{12pt}\addtocounter{sectionc}{1} 
\setcounter{subsectionc}{0}\setcounter{subsubsectionc}{0}\noindent 
        {\tenbf\thesectionc. #1}\par\vspace{5pt}}
\renewcommand{\subsection}[1] {\vspace{12pt}\addtocounter{subsectionc}{1} 
        \setcounter{subsubsectionc}{0}\noindent 
        {\bf\thesectionc.\thesubsectionc. {\kern1pt \bfit #1}}\par\vspace{5pt}}
\renewcommand{\subsubsection}[1] {\vspace{12pt}\addtocounter{subsubsectionc}{1}
        \noindent{\tenrm\thesectionc.\thesubsectionc.\thesubsubsectionc.
        {\kern1pt \tenit #1}}\par\vspace{5pt}}
\newcommand{\nonumsection}[1] {\vspace{12pt}\noindent{\tenbf #1}
        \par\vspace{5pt}}

\newcounter{appendixc}
\newcounter{subappendixc}[appendixc]
\newcounter{subsubappendixc}[subappendixc]
\renewcommand{\thesubappendixc}{\Alph{appendixc}.\arabic{subappendixc}}
\renewcommand{\thesubsubappendixc}
        {\Alph{appendixc}.\arabic{subappendixc}.\arabic{subsubappendixc}}

\renewcommand{\appendix}[1] {\vspace{12pt}
        \refstepcounter{appendixc}
        \setcounter{figure}{0}
        \setcounter{table}{0}
        \setcounter{lemma}{0}
        \setcounter{theorem}{0}
        \setcounter{corollary}{0}
        \setcounter{definition}{0}
        \setcounter{equation}{0}
        \renewcommand{\thefigure}{\Alph{appendixc}.\arabic{figure}}
        \renewcommand{\thetable}{\Alph{appendixc}.\arabic{table}}
        \renewcommand{\theappendixc}{\Alph{appendixc}}
        \renewcommand{\thelemma}{\Alph{appendixc}.\arabic{lemma}}
        \renewcommand{\thetheorem}{\Alph{appendixc}.\arabic{theorem}}
        \renewcommand{\thedefinition}{\Alph{appendixc}.\arabic{definition}}
        \renewcommand{\thecorollary}{\Alph{appendixc}.\arabic{corollary}}
        \renewcommand{\theequation}{\Alph{appendixc}.\arabic{equation}}
        \noindent{\tenbf Appendix \theappendixc #1}\par\vspace{5pt}}
\newcommand{\subappendix}[1] {\vspace{12pt}
        \refstepcounter{subappendixc}
        \noindent{\bf Appendix \thesubappendixc. {\kern1pt \bfit #1}}
        \par\vspace{5pt}}
\newcommand{\subsubappendix}[1] {\vspace{12pt}
        \refstepcounter{subsubappendixc}
        \noindent{\rm Appendix \thesubsubappendixc. {\kern1pt \tenit #1}}
        \par\vspace{5pt}}

\newcommand{\textlineskip}{\baselineskip=13pt}
\newcommand{\smalllineskip}{\baselineskip=10pt}

\def\eightcirc{
\begin{picture}(0,0)
\put(4.4,1.8){\circle{6.5}}
\end{picture}}
\def\eightcopyright{\eightcirc\kern2.7pt\hbox{\eightrm c}} 

\newcommand{\copyrightheading}[1]
        {\vspace*{-2.5cm}\smalllineskip{\flushleft
        {\footnotesize International Journal of Modern Physics A, #1}\\
        {\footnotesize $\eightcopyright$\, World Scientific Publishing
         Company}\\
         }}

\def\abstracts#1#2#3{{
        \centering{\begin{minipage}{4.5in}\baselineskip=10pt\footnotesize
        \parindent=0pt #1\par 
        \parindent=15pt #2\par
        \parindent=15pt #3
        \end{minipage}}\par}} 

\renewenvironment{thebibliography}[1]
        {\frenchspacing
         \ninerm\baselineskip=11pt
         \begin{list}{\arabic{enumi}.}
        {\usecounter{enumi}\setlength{\parsep}{0pt}
         \setlength{\leftmargin 12.7pt}{\rightmargin 0pt} %FOR 1--9 ITEMS
         \setlength{\itemsep}{0pt} \settowidth
        {\labelwidth}{#1.}\sloppy}}{\end{list}}

\newcounter{itemlistc}
\newcounter{romanlistc}
\newcounter{alphlistc}
\newcounter{arabiclistc}

\newcommand{\fcaption}[1]{
        \refstepcounter{figure}
        \setbox\@tempboxa = \hbox{\footnotesize Fig.~\thefigure. #1}
        \ifdim \wd\@tempboxa > 5in
           {\begin{center}
        \parbox{5in}{\footnotesize\smalllineskip Fig.~\thefigure. #1}
            \end{center}}
        \else
             {\begin{center}
             {\footnotesize Fig.~\thefigure. #1}
              \end{center}}
        \fi}

\newcommand{\tcaption}[1]{
        \refstepcounter{table}
        \setbox\@tempboxa = \hbox{\footnotesize Table~\thetable. #1}
        \ifdim \wd\@tempboxa > 5in
           {\begin{center}
        \parbox{5in}{\footnotesize\smalllineskip Table~\thetable. #1}
            \end{center}}
        \else
             {\begin{center}
             {\footnotesize Table~\thetable. #1}
              \end{center}}
        \fi}

\def\@citex[#1]#2{\if@filesw\immediate\write\@auxout
        {\string\citation{#2}}\fi
\def\@citea{}\@cite{\@for\@citeb:=#2\do
        {\@citea\def\@citea{,}\@ifundefined
        {b@\@citeb}{{\bf ?}\@warning
        {Citation `\@citeb' on page \thepage \space undefined}}
        {\csname b@\@citeb\endcsname}}}{#1}}

\newif\if@cghi
\def\cite{\@cghitrue\@ifnextchar [{\@tempswatrue
        \@citex}{\@tempswafalse\@citex[]}}
\def\citelow{\@cghifalse\@ifnextchar [{\@tempswatrue
        \@citex}{\@tempswafalse\@citex[]}}
\def\@cite#1#2{{$\null^{#1}$\if@tempswa\typeout
        {IJCGA warning: optional citation argument 
        ignored: `#2'} \fi}}

\def\pmb#1{\setbox0=\hbox{#1}
        \kern-.025em\copy0\kern-\wd0
        \kern.05em\copy0\kern-\wd0
        \kern-.025em\raise.0433em\box0}

\def\fnt#1#2{\footnotetext{\kern-.3em
        {$^{\mbox{\scriptsize #1}}$}{#2}}}

\def\fpage#1{\begingroup
\voffset=.3in
\thispagestyle{empty}\begin{table}[b]\centerline{\footnotesize #1}
        \end{table}\endgroup}

\def\runninghead#1#2{\pagestyle{myheadings}
\markboth{{\protect\footnotesize\it{\quad #1}}\hfill}
{\hfill{\protect\footnotesize\it{#2\quad}}}}
\font\tenrm=cmr10
\font\tenit=cmti10 
\font\tenbf=cmbx10
\font\bfit=cmbxti10 at 10pt
\font\ninerm=cmr9

\font\eightrm=cmr8

\def\qed{\hbox{${\vcenter{\vbox{                        %HOLLOW SQUARE
   \hrule height 0.4pt\hbox{\vrule width 0.4pt height 6pt
   \kern5pt\vrule width 0.4pt}\hrule height 0.4pt}}}$}}

\begin{document}
\runninghead{Awad and Johnson, Scale Invariance and the AdS/CFT
  Correspondence} {Awad  and Johnson, Scale Invariance and the AdS/CFT
  Correspondence} \normalsize\textlineskip \thispagestyle{empty}
\setcounter{page}{1}

\copyrightheading{}   %{Vol. 0, No. 0 (1993) 000--000}

\vspace*{0.88truein}

\fpage{1} \centerline{\bf Scale Invariance and the AdS/CFT
  Correspondence} \vspace*{0.37truein} \centerline{\footnotesize ADEL
  M. AWAD\footnote{adel@pa.uky.edu} \footnote{The speaker in DPF
    2000}} \vspace*{0.015truein} \centerline{\footnotesize\it
  Department of Physics and Astronomy, University of Kentucky}
\baselineskip=10pt \centerline{\footnotesize\it Lexington, Kentucky
  40506, U.S.A.} \vspace*{10pt} \centerline{\footnotesize
  CLIFFORD V. JOHNSON\footnote{c.v.johnson@durham.ac.uk}}
\vspace*{0.015truein} \centerline{\footnotesize\it Centre for Particle
  Theory, Department of Mathematical Sciences} \baselineskip=10pt
\centerline{\footnotesize\it University of Durham, Durham DH1 3LE,
  England, U.K.}
%\vspace*{0.225truein}
%\publisher{(received date)}{(revised date)}
\vspace*{0.21truein} \abstracts{Using the AdS/CFT correspondence, we
  show that the Anti-de Sitter (AdS) rotating (Kerr) black holes in
  five and seven dimensions provide us with examples of non--trivial
  field theories which are scale, but not conformally invariant. This
  is demonstrated by our computation of the actions and the
  stress--energy tensors of the four and six dimensional field
  theories residing on the boundary of these Kerr--AdS black holes
  spacetimes.}{}{}

\textlineskip   %) USE THIS MEASUREMENT WHEN THERE IS
\vspace*{12pt}   %) NO SECTION HEADING

In situations when a non--trivial field theory is exactly scale
invariant, it often follows that it is also conformally invariant.
This is sometimes regarded as a rule of thumb, although it is known to
be not generally applicable. Discussion, and few counterexamples may
be found in the literature\cite{jackiwone}$\!\!^{-}$\cite{ivo}.  These
counterexamples, all in flat spacetime, are not generic. Here, we
construct more counterexamples by placing the field theory in curved
spacetime $\cal M$, and exploit the general structure of the conformal
anomaly. The general form of the anomaly in dimension $n$ (which is
even, since we only have conformal anomalies in those cases) is given
by\cite{deser}:
\begin{equation} 
{\widehat T}_a^a= {\rm c}_0 E_n + \sum_i {\rm c}_i I_i + \nabla_a J^a\ . 
\end{equation}
Here, the ${\rm c}$'s are constants, $E_n$ is the Euler density, $I_i$
are terms constructed from the Weyl tensor and its derivatives, and
the last term is a collection of total derivative terms. The first
type of term is called ``type~A'', the next ``type~B'', and the last
``type~D''.  We note that non--trivial counterexamples can be
generated by considering a theory with conformal anomaly, which
satisfies $\int_{\cal M} d^nx \sqrt{\gamma}\, {\widehat T}^a_a=0$,
preserving scale invariance. This can be done by choosing a
non--vanishing type A anomaly and a vanishing type B anomaly. The type
D anomaly always can be removed by adding local counterterms. We now
need to find a prescription for defining (and studying the properties
of) a sensible theory on a spacetime with properties chosen to yield
the desired anomalies, producing thus our counterexamples.

In this work, we show how to do this. We use the tools which were
developed in the context of the ``AdS/CFT
correspondence''\cite{juan,witten,gubkleb} in our analysis. One of
these tools is the boundary counterterm method\cite{Balasubramanian}
which we use to calculate the quasi local stress--energy tensor of an
AdS gravitational solution (see also refs.[14,15]). According to the
correspondence, this stress-energy tensor, evaluated on the boundary
at infinity, is equal to the expectation value of that of a
renormalized field theory (at strong coupling) in one dimension fewer.
The field theory (see refs.[7--9] for details of exactly which field
theory, depending upon dimension) is conformal in flat spacetime, but
in general lives on a spacetime which shares a metric with the
boundary of the AdS solution.  Complete details of our computations of
the action and stress tensor components may be found in the
literature\cite{adel,svc,hd}, and we present only a small part of the
results here.

Our examples come from the Kerr--AdS spacetimes in five and seven
dimensions (with only one of the two rotation parameters, which we
shall call $a$, non--zero\cite{Hawkingtwo}):
\begin{eqnarray} ds^2&=&-{\Delta_{r} \over {\rho}^2}\left(dt-{a
      \sin^2{\theta} \over \Xi } d\phi\right)^2 +r^2 \cos^2{\theta}
  d\psi^2+{{\rho}^2\over\Delta_\theta}d\theta^2+{\rho^2 \over
    \Delta_{r}}dr^2\nonumber\\
  & &+{\Delta_{\theta}\sin^2{\theta}\over\rho^2}\left(adt-{(r^2+a^2)
      \over \Xi} d\phi\right)^2+r^2 \cos^2{\theta} d \Omega_{n-3}^2\ ,
\label{metricstuff}
\end{eqnarray}
where $d \Omega_{n-3}^2$ is the unit metric on $S^{n-3}$, and 
\begin{eqnarray}
\rho^2 &=& r^2+a^2\cos^2(\theta)\ ,\quad \Xi=1-a^2/l^2,\nonumber\\
     \Delta_{r}&=&(r^2+a^2)(1+r^2/l^2)-2MG/r\ ,\nonumber\\
\Delta_{\theta}&=&1-a^2/l^2 \cos^2(\theta)\ .
\end{eqnarray}
The dimension of the gravitational solution is $n+1$, while that of
the field theory is $n$.  The metric, $\gamma_{ab}$, on which the
field theory (either the ${\cal N}{=}4$ Yang--Mills theory for $n{=}4$
or the (0,2) supersymmetric conformal field theory (see ref.[10] for a
review) for $n{=}6$) resides is that of a rotating Einstein universe,
with line element\cite{Hawkingtwo}:
\begin{equation}
ds^2=-dt^2+{2a\sin^2{\theta}\over\Xi}dtd\phi+l^2{d\theta^2 \over
\Delta_{\theta}}+l^2{\sin^2{\theta} \over
\Xi}d\phi^2+l^2\cos^2{\theta} d \Omega_{n-3}^2\ .
\label{rotatestein}
\end{equation}

The stress tensor for strongly coupled ${\cal N}{=}4$ Yang--Mills on
this spacetime ({\it i.e.,} $n=4$) computed from the Kerr--AdS$_5$
spacetime was computed in ref.\cite{adel} The trace is:
\begin{equation} 
{\widehat T}_{a}^{a}=-{N^2a^2 \over 4 \pi^2l^6}[a^2/l^2 
(3\cos^4{\theta}-2\cos^2{\theta})-\cos{2\theta}]\ .
\end{equation}
\\
It is non--zero, and so conformal invariance is broken.  A quick
computation shows that this is a total derivative. In fact, ${\widehat
  T}_{a}^{a}={-}(N^2/\pi^2)E_4$, where $E_{4}$ is Euler density in
four dimensions\cite{svc}, but the topology is trivial.  The
coefficient is precisely the field theory value\cite{Henningson}.
Since~$\int d^4\!x \sqrt{-\gamma}\, {\widehat T}_{a}^{a}{=}0$, the
theory is scale invariant. Note that the Euler density is proportional
to $\Box{\cal R}$. It therefore appears to be of type~D, and one might
imagine\cite{adel,ho} that it can be removed by a local counterterm.
However, in general it cannot be written in this form, and this is a
red herring. Our anomaly is purely of type A, and as such its
coefficient cannot be changed by adding local counterterms. Instead,
we must accept the presence of the anomaly and give up conformal
invariance; the rotation parameter $a$ (which simply corresponds to
the inclusion in the field theory of a chemical potential coupling to
angular velocity\cite{Hawkingthree}) has broken conformal invariance
of the theory, but scale invariance is preserved.

Consider now the case of Kerr--AdS$_{7}$.  We computed\cite{svc,hd}
the non--vanishing components for the stress tensor at large $r$. The
trace of the stress tensor yields:
\begin{eqnarray} 
{\widehat T}_a^a&=&-{a^2 N^3 \over 2\pi^3 l^8
    }\left[5a^4/l^4 \cos^6\theta -8 \cos^4\theta
    a^2/l^2(1+a^2/l^2)-2(1+a^2/l^2)\right.\nonumber\\ &
  &\left.+3\cos^2\theta(1+a^4/l^4+3a^2/l^2)\right]\ .
\end{eqnarray}

We find that\cite{svc} ${\widehat T}_a^a={-}(N^3 / 4508 \pi^3)E_6$,
where $E_{6}$ is the Euler density in six dimension\cite{tseytlin}.
The coefficient matches the field theory results in the
literature\cite{Hawkingtwo,tseytlin,kraus}.  Again, since the topology
is trivial, $\int d^6x \sqrt{\gamma} \,{\widehat T}_a^a=0 $,
preserving scale invariance but not conformal invariance, thus
completing our second counterexample.

\nonumsection{References}

{}

\noindent

\begin{thebibliography}{99999}
\bibitem{jackiwone}
C.G.~Callan, S.~Coleman and R.~Jackiw,
%``A New Improved Energy - Momentum Tensor,''
Ann.\ Phys.\ {\bf 59}, 42 (1970).

\bibitem{jackiwtwo}S.~Coleman and R.~Jackiw,
%``Why Dilatation Generators Do Not Generate Dilatations?,''
Annals Phys.\  {\bf 67}, 552 (1971).

\bibitem{paul}C.~M.~Hull and P.~K.~Townsend,
Nucl.\ Phys.\ {\bf B274}, 349 (1986).

\bibitem{joe}J.~Polchinski,
%``Scale And Conformal Invariance In Quantum Field Theory,''
Nucl.\ Phys.\ {\bf B303}, 226 (1988).

\bibitem{ivo}A. Iorio, L. O'Raifeartaigh, I. Sachs,
 C. Wiesendanger, Nucl. Phys. {\bf B495} (1997) 433, hep-th/9607110.

\bibitem{deser}S.~Deser and A.~Schwimmer,
%``Geometric classification of conformal anomalies in arbitrary dimensions,''
Phys.\ Lett.\  {\bf B309} (1993) 279, hep-th/9302047.

\bibitem{juan}J. Maldacena, Adv. Theor. Math. Phys. {\bf 2} 231 (1998),
hep-th/9711200.


\bibitem{witten}E. Witten, Adv. Theor. Math.
Phys. {\bf 2}  253 (1998), hep-th/9802150.
 
\bibitem{gubkleb}S. S. Gubser, I. R. Klebanov and A. M. Polyakov,
  Phys. Lett.  {\bf B428}105 (1998), hep-th/9802109.


\bibitem{agmoo}
O.~Aharony, S.~S.~Gubser, J.~Maldacena, H.~Ooguri and Y.~Oz,
%``Large N field theories, string theory and gravity,''
Phys.\ Rept.\  {\bf 323}, 183 (2000), hep-th/9905111.

\bibitem{Balasubramanian}V.~Balasubramanian and P.~Kraus,
%``A stress tensor for anti-de Sitter gravity,''
Commun.\ Math.\ Phys.\  {\bf 208}, 413 (1999), hep-th/9902121.
\bibitem{Brown}J. D. Brown and J. W. York, Phys. Rev. {\bf D47}, 1407 (1993).

\bibitem{counter}R.~Emparan, C.~V.~Johnson and R.~C.~Myers,
%``Surface terms as counterterms in the AdS/CFT correspondence,''
Phys.\ Rev.\ {\bf D60}, 104001 (1999), hep-th/9903238.

\bibitem{Henningson}M. Henningson and K. Skenderis, J.H.E.P.
9807 023 (1998), hep-th/9806087.

\bibitem{adel}A. Awad and C. V. Johnson, Phys. Rev. {\bf D61} (2000)
084025, hep-th/9910040.

\bibitem{svc}A. Awad and C. V. Johnson, to appear in Phys. Rev. D,
  hep-th/0006037.
  
\bibitem{hd}Adel M. Awad and C. V. Johnson, to appear in Phys. Rev. D,
  hep-th/0008211.




\bibitem{Hawkingtwo}S. W. Hawking, C. J. Hunter and M. M. Taylor--Robinson,
Phys. Rev. {\bf D59} (1999) 064005, hep-th/9811056.


\bibitem{ho}J.~Ho, hep-th/9910124 and hep-th/0005250.

\bibitem{Hawkingthree}  S. W. Hawking and  H. S. Reall,
  Phys. Rev. {\bf D61} (2000) 024014, hep-th/9908109.
\bibitem{tseytlin}F.~Bastianelli, S.~Frolov and A.~A.~Tseytlin,
%``Conformal anomaly of (2,0) tensor multiplet in six 
%dimensions and  AdS/CFT correspondence,''
JHEP {\bf 0002}, 013 (2000), hep-th/0001041.

\bibitem{kraus}P. Kraus, F Larsen, and R. Siebelink, Nucl. Phys. {\bf B563}, 259 (1999).

\end{thebibliography}
\end{document}